\documentclass[aps,prl,twocolumn,superscriptaddress,groupedaddress]{revtex4}  
\usepackage{graphicx}  
\usepackage{dcolumn}   
\usepackage{bm}        
\usepackage{amssymb}   
\usepackage[dvipsnames]{xcolor}
\begin{document}

\title{Ferromagnetic resonance assisted optomechanical magnetometer}

\author{M. F. Colombano}
\affiliation{Catalan Institute of Nanoscience and Nanotechnology (ICN2), CSIC and BIST, Campus UAB, Bellaterra, 08193 Barcelona, Spain}
\affiliation{Depto. de F\`isica, Universitat Aut\`onoma de Barcelona, 08193 Bellaterra, Spain}

\author{G. Arregui}
\affiliation{Catalan Institute of Nanoscience and Nanotechnology (ICN2), CSIC and BIST, Campus UAB, Bellaterra, 08193 Barcelona, Spain}
\affiliation{Depto. de F\`isica, Universitat Aut\`onoma de Barcelona, 08193 Bellaterra, Spain}

\author{F. Bonell}
\affiliation{Catalan Institute of Nanoscience and Nanotechnology (ICN2), CSIC and BIST, Campus UAB, Bellaterra, 08193 Barcelona, Spain}

\author{N. E. Capuj}
\affiliation{Depto. F\'{i}sica, Universidad de La Laguna, 38200 San Crist\'{o}bal de La Laguna, Spain}
\affiliation{Instituto Universitario de Materiales y Nanotecnolog\'{i}a, Universidad de La Laguna, 38071 Santa Cruz de Tenerife, Spain.}

\author{E. Chavez-Angel}
\affiliation{Catalan Institute of Nanoscience and Nanotechnology (ICN2), CSIC and BIST, Campus UAB, Bellaterra, 08193 Barcelona, Spain}


\author{A. Pitanti}
\affiliation{NEST, CNR - Istituto Nanoscienze and Scuola Normale Superiore, Piazza San Silvestro 12, 56127 Pisa, Italy}

\author{S.O. Valenzuela}
\affiliation{Catalan Institute of Nanoscience and Nanotechnology (ICN2), CSIC and BIST, Campus UAB, Bellaterra, 08193 Barcelona, Spain}
\affiliation{ICREA - Instituci\`o Catalana de Recerca i Estudis Avan\c{c}ats, 08010 Barcelona, Spain}

\author{C. M. Sotomayor-Torres}
\affiliation{Catalan Institute of Nanoscience and Nanotechnology (ICN2), CSIC and BIST, Campus UAB, Bellaterra, 08193 Barcelona, Spain}
\affiliation{ICREA - Instituci\`o Catalana de Recerca i Estudis Avan\c{c}ats, 08010 Barcelona, Spain}

\author{D. Navarro-Urrios}
\altaffiliation{Corresponding author}
\email{dnavarro@ub.edu}
\affiliation{MIND-IN2UB, Departament d'Enginyer\`ia Electr\`onica i Biom\`edica, Facultat de F\`isica, Universitat de Barcelona, Mart\`i i Franqu\`es 1, 08028 Barcelona, Spain}

\author{M. V. Costache}
\affiliation{Catalan Institute of Nanoscience and Nanotechnology (ICN2), CSIC and BIST, Campus UAB, Bellaterra, 08193 Barcelona, Spain}

\date{\today}

\begin{abstract}
The resonant enhancement of mechanical and optical interaction in optomechanical cavities enables their use as extremely sensitive displacement and force detectors. In this work we demonstrate a hybrid magnetometer that exploits the coupling between the resonant excitation of spin waves in a ferromagnetic insulator and the resonant excitation of the breathing mechanical modes of a glass microsphere deposited on top. The interaction is mediated by magnetostriction in the ferromagnetic material and the consequent mechanical driving of the microsphere. The magnetometer response thus relies on the spectral overlap between the ferromagnetic resonance and the mechanical modes of the sphere, leading to a peak sensitivity of 850 pT Hz$^{-1/2}$ at 206 MHz when the overlap is maximized. By externally tuning the ferromagnetic resonance frequency with a static magnetic field we demonstrate sensitivity values at resonance around a few nT Hz$^{-1/2}$ up to the GHz range. Our results show that our hybrid system can be used to build a high-speed sensor of oscillating magnetic fields.
\end{abstract}

\maketitle



Cavity optomechanics (OM) focuses on the low-energy interaction between photons and micro and nanomechanical systems embedded in an optical cavity with main applications as high performance detectors and as interfaces for quantum information processing \cite{pirkkalainen2015cavity,LIGO,kippenberg2008,forcesensor,accelerometer}. OM cavities enable ultra-sensitive optical transduction of mechanical motion in km-scale systems such as LIGO \cite{LIGO} down to nano-scale quantum resonators \cite{kippenberg2008}. Stimuli read-out experiments based on such platforms have already reached the state-of-the-art for force sensors \cite{forcesensor} and accelerometers \cite{accelerometer}. In addition, the interaction of mechanical elements with magnetic fields also makes OM devices high performance magnetometers, i.e,  room temperature OM magnetometer (OMM) of small size \cite{bowen, onchipmagnetometer}, high sensitivity \cite{quantummag} and large dynamic range \cite{fostneradv}. The ability to measure small magnetic fields over a broad frequency range is important for numerous applications playing a key role in areas such as geology \cite{meyer2005squid}, space exploration \cite{bookmagnetometers}, biology \cite{edelstein} and medical imaging \cite{glenn2018high}.

In this Letter, we report a mechanical and magnetic hybrid resonator based on a thin film of magnetic insulator yttrium iron garnet (YIG) coupled to a glass microsphere optical cavity. We show that the combined coupling of optical whispering gallery modes (WGM) to mechanical breathing modes in the microsphere and the presence of a ferromagnetic resonance (FMR) in the YIG film enables a sensitive  RF magnetic-field detector. When the magnetic field frequency is able to excite the YIG FMR and further coincides with a mechanical mode of the sphere, the detection sensitivity is maximized. The basic transduction principle involves the conversion of an RF magnetic field, that resonantly excites magnons, into mechanical vibrations via magnetostriction in the YIG film \cite{Litvinenko2015, polzikova2013,polzikova2016, Litvinenko2020}. Although the frequencies of the FMR mode and mechanical breathing mode in the microsphere may differ, the FMR frequency can be tuned by a static magnetic field until both resonances are aligned, hence giving rise to a modulation of the microsphere WGM due to the OM interaction. We show that, by following that procedure, it is possible to maximize the peak sensitivity at the mechanical mode frequencies. This allows the magnetometer to operate in multiple windows where mechanical modes are found, from 50 MHz to 1.1GHz. The sensitivities obtained for those frequencies are in the nT Hz $^{-1/2}$ range. We attain a maximum sensitivity of $\sim$ 850 pT Hz$^{-1/2}$ at 206 MHz operating at room temperature.

\begin{figure*}[t!]
 \centering
\includegraphics[width=\textwidth]{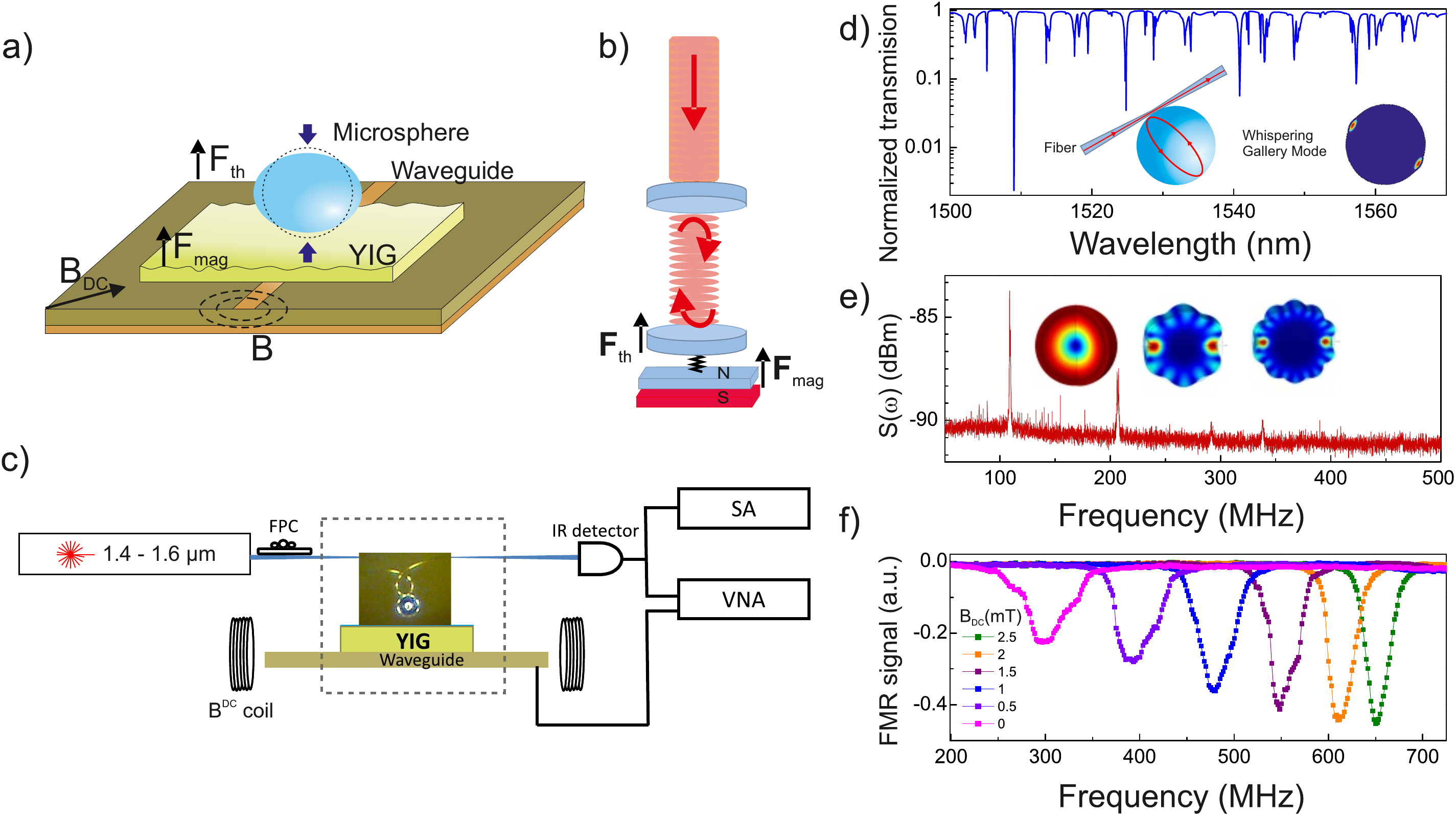}
      \caption{(\textbf{a}) Schematic representation of the magnetometer, including a microstrip waveguide, a ferromagnetic YIG film and a barium-titanium-silicate microsphere.  (\textbf{b}) Conceptual schematic of the device. The optomechanical system is modeled as a Fabry-P\'erot cavity with a moving mirror. \textbf{F$_{th}$} and \textbf{F$_{mag}$} denote the thermal force and the magnetostrictive forces, respectively. (\textbf{c}) A simplified schematic of the full experimental setup. A tapered fiber is used to probe the optical modes of the microsphere. TE polarization is set with a fiber polarization controller (FPC) and the transmitted signal is sent to a fast photodetector. Two electromagnetic coils are used to generate static magnetic fields that tune ferromagnetic resonance modes on the YIG film. (\textbf{d}) Optical whispering gallery modes spectrum measured on the microsphere. The inset shows a schematic of the coupling scheme between the tapered and the sphere and a simulation of the optical WGMs. (\textbf{e}) Mechanical mode spectrum of the microsphere. Only thermally driven motion is observed corresponding to radial breathing modes. The inset shows the displacement profile of the first three modes. (\textbf{f}) FMR resonances of the YIG film applying different static magnetic fields measured by detecting the reflected signal $Re(S_{11})$.}
\label{fig1}
\end{figure*}

An illustration of the main part of the hybrid system is shown in Fig. \ref{fig1}a. It consists of a 0.5$\times$0.5 mm$^{2}$  and 1 $\mu$m thick film of YIG grown over a Gadolinium Garnet substrate. Glass microspheres of Barium-Titanium-Silicate (BTS) with a diameter between 40 $\mu$m and 70 $\mu$m \cite{Urrios2015,Martin2013} were deposited on the YIG thin film. These microspheres are used as high-quality OM cavities supporting both optical WGM and mechanical breathing modes with large  OM coupling (G) values, defined as the optical frequency shift by a unit mechanical displacement \cite{aspelmeyer2014}. A schematic picture of the OMM principle is shown in Fig. \ref{fig1}b. The sensor can be modeled as a Fabry-Perot optical interferometer in which one of the mirrors responds mechanically to an applied magnetic field. This response is due to magnetostriction in the bulk of the material, i.e, the generation of an oscillating stress when a magnetic field is applied. This stress acts as a source force for mechanical motion of the mirror, greatly amplified when the initial drive is resonant with a mechanical eigenfrequency.

In Fig. \ref{fig1}b, we also illustrate that the two main forces actuating the mechanical modes are the thermal Langevin force ($F_{th}$), and the one associated with the RF magnetic field ($F_{mag}$). These forces cause a variation of the cavity length, $x$, shifting the optical resonance by $\delta\omega_{optical}=Gx$.

The experimental setup is shown in Fig. \ref{fig1}c. An infrared tunable laser source is used to couple light into the WGM of the microsphere using a micro-looped tapered fiber placed close enough to ensure an overlap between the evanescent field of the fiber fundamental mode and the WGM. The mechanical motion is detected by measuring the RF modulation of the transmitted light, which is collected and sent to a photo-detector with an operational bandwidth of 12 GHz. The output signal can be analyzed by a spectrum analyzer (SA) and a vector network analyzer (VNA). The latter is also used to inject RF signals into a shorted end microstrip waveguide (MSW), which we use to generate RF magnetic fields (see Fig. \ref{fig1}c, and Supplementary Information), and to characterize the FMR modes of the YIG film. A static magnetic field B$_{DC}$ is generated by two coils connected to a current source. All experiments were carried out under ambient conditions of temperature and pressure.

In Fig. \ref{fig1}d we show the typical optical transmission spectrum from a microsphere of about 40 $\mu$m in diameter with multiple WGMs resonances. The mode used here near 1509 nm has a quality factor of 10$^{8}$ and couples to several mechanical radial breathing modes (Fig. \ref{fig1}e), at 109, 206, 292, 338, and 465 MHz. The displacement profiles of the first three modes obtained using COMSOL Multiphysics software are shown in Fig. \ref{fig1}e (inset). Due to the disadvantageous refractive index contrast between the YIG film ($n =$ 2.19) and the BTS microsphere sphere ($n =$ 1.9), we use an intermediate 100 nm thick layer of Polymethyl methacrylate (PMMA), which has a smaller refractive index ($n =$ 1.49), to preserve the high Q-factors of the optical modes. 

The method used to excite and measure uniform magnon modes, from MHz to GHz range, in the YIG film, is the broadband FMR method. The MSW creates a RF field perpendicular to B$_{DC}$ that excites the precessional motion of the magnetization around B$_{DC}$. When the excitation frequency matches the FMR condition, energy is absorbed. Consequently, a dip is observed in the reflection spectrum (S$_{11}$). With the magnetic field generated in our set up (B$_{DC}$ $\leq$ 10 mT), the FMR frequency can be tuned from 0.2 to 1 GHz (see Fig. \ref{fig1}f). In the case of a thin film with a static magnetic field applied in-plane and the RF field perpendicular to the direction of magnetization M, the resonance frequency is $\omega_{0}\equiv\gamma\sqrt{B_{int}(B_{int}+\mu M)}$, with the internal field $B_{int}=B_{an}+$B$_{DC}$, $B_{an}$ being the anisotropy field and $\gamma$ the gyromagnetic ratio (See Supplementary Information). We note that the linewidth of the FMR mode (Fig. \ref{fig1}f) decreases as a function of B$_{DC}$ from about 46 MHz to 24 MHz (at 2.5 mT). At low B$_{DC}$ values, the magnetization is not uniform, leading to inhomogeneous spectral broadening of the resonance.



\begin{figure}[t!]
 \centering
  \includegraphics[width=8.9cm]{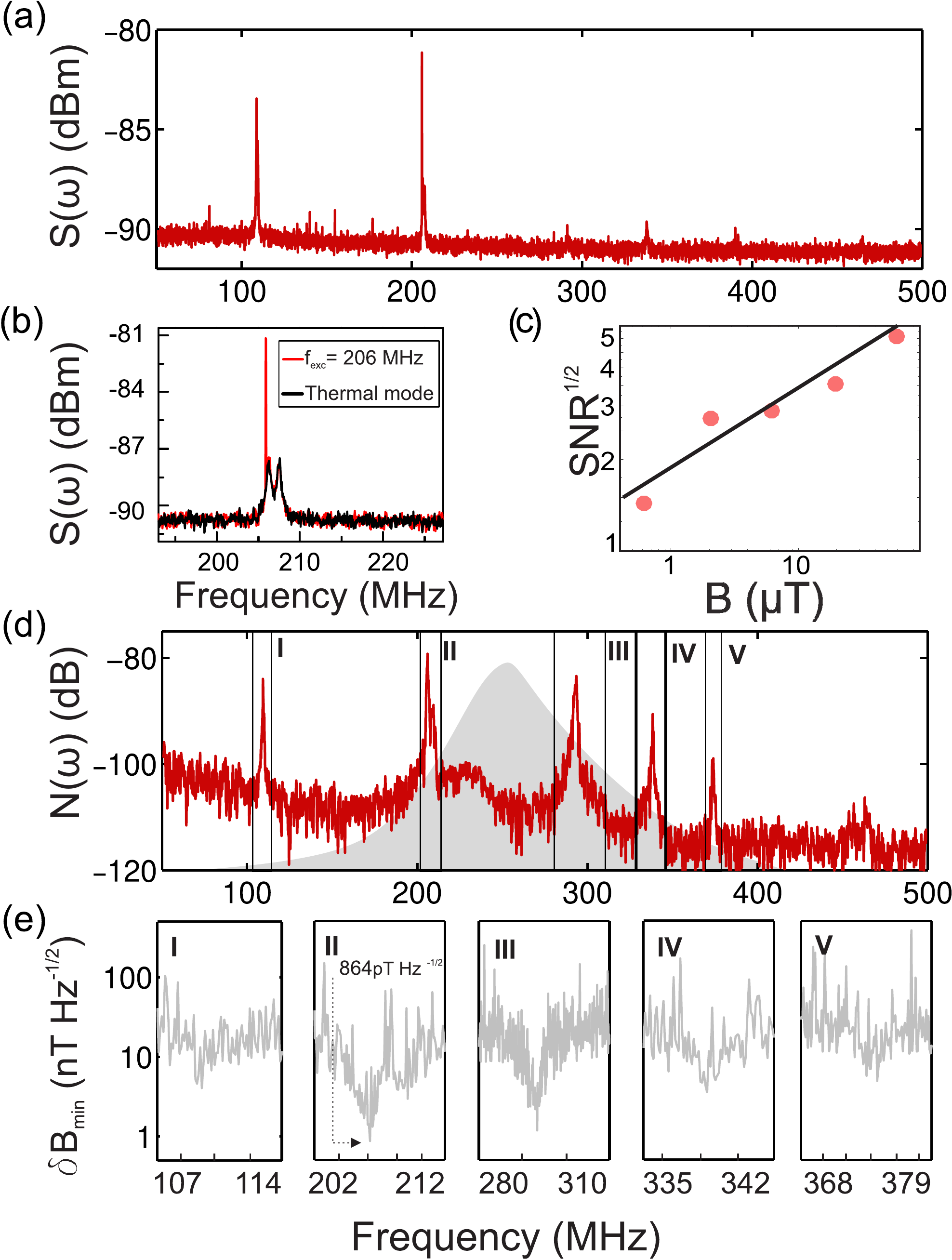}
      \caption{(\textbf{a}) Spectral response of the magnetometer  $S(\omega )$ measured with a spectrum analyzer. (\textbf{b}) Mechanical spectrum of the  microsphere excited by applying an RF magnetic field of 2.3 $\mu$T at 206 MHz, in red. The black curve corresponds to the mechanical mode without excitation. (\textbf{c}) Square root of signal to noise ratio (SNR) of the system as a function of the applied RF magnetic field. (\textbf{d}) System response $N(\omega)$ as a function of the frequency of the RF field. The FMR is shown (inverted) on the back of the graph to illustrate the mechanical modes affected by the resonant effect. (\textbf{e}) Magnetic field sensitivity $\delta B_{min}(\omega)$ as a function of frequency defined where there is overlapped with mechanical modes. Five excited mechanical modes (labelled from I to V) allow to calculate the sensitivity in a frequency window of $\thicksim$ 10 MHz around the frequencies of the mechanical resonances. A peak sensitivity of $\delta B_{min}(\omega)\sim$ 850 pT Hz$^{-1/2}$ is achieved.}
\label{fig6}
\end{figure}

The spectral response of the system to an applied magnetic excitation at a calibration frequency, $\omega_{cal}$, is shown in Fig. \ref{fig6}a. In Fig. \ref{fig6}b, we plot a zoom area with and without excitation. The thermal spectrum (black line) shows a double-peak response between 205 and 210 MHz. When the system is excited at a calibration frequency $\omega_{cal}$ = 206 MHz, with an RF power level of -10 dBm (2.3 $\mu T$), a sharp peak emerges (red line). Such a peak disappears when the sphere is lifted from the YIG and the mechanical contact is lost, which demonstrates the mechanical origin of the signal. The force induced by vibrations in the YIG modifies the mechanical spectrum of the microsphere with a corresponding Signal-to-Noise-Ratio (SNR) of 8 dB. As shown in Fig. \ref{fig6}c, we observe a linear dependence of square root of SNR at $\omega_{cal}$ on the applied RF magnetic field magnitude ($B$). Here $B$ is estimated using the characteristic impedance of the of circuit and the applied RF power (see Supplementary Information). The $B$ sensitivity ($B_{min}$) is given by the field strength at which the spectral peak height is equal to the noise (SNR=1) for a 1 Hz measurement resolution bandwidth (RBW) \cite{fostner}. The corresponding magnetic field at $\omega_{cal}$ is $B_{min}(\omega_{cal})$ = 0.5 $\mu$T for RBW = 30 kHz. Then, the sensitivity at $\omega_{cal}$ is given by $\delta B_{min}(\omega_{cal}) = \frac{B_{min}(\omega_{cal})}{\sqrt{RBW}} \sim$ 3 nT Hz$^{-1/2}$. The dynamic response of the sensor, N($\omega$), over a wide frequency range is obtained by varying the input frequency from port 1 of the VNA and by looking at S$_{21}$, where port 2 is directly connected to the detector. As shown in Fig. \ref{fig6}d, we observe a peak in signal N($\omega$) wherever $\omega$ is resonant with a mechanical mode (labelled in Fig. \ref{fig6}d from I to IV) with a high OM coupling (see Fig. \ref{fig1}e). Due to the enhanced noise rejection of the VNA, we can detect modes at 374 MHz and 456 MHz that in the thermally activated spectrum (Fig. \ref{fig6}c) were below the noise level. By following a similar procedure as in Ref.\cite{bowen}, the frequency dependence of the sensitivity $\delta B_{min}(\omega)$ is obtained by combining the spectral calibration  at a single frequency $\omega_{cal}$, the noise power spectrum in absence of a magnetic field $S(\omega)$, and $N(\omega)$ on the mechanical modes.

\begin{equation}
\label{eq:eq1}
\delta B_{min}(\omega)=\sqrt{\frac{S(\omega) N(\omega_{cal})}{N(\omega) S(\omega_{cal})}}\delta B_{min}(\omega_{cal})
\end{equation}

Fig. \ref{fig6}e plots the sensitivity within the frequency range associated to the first five mechanical modes of the microsphere observed in Fig. \ref{fig1}c. The lowest sensitivity value obtained is $\sim$ 850 pT Hz$^{-1/2}$ close to the mechanical mode at 206 MHz. This particular mode presents a large overlap with the optical WGM, since its displacement field profile is concentrated along the edge of the sphere (see Fig. \ref{fig1}a and \ref{fig1}e). It is also worth noting that the frequency of the mode is still several tens of MHz away from the center of the FMR resonance (see Fig. \ref{fig6}d) so that the reported value of minimum sensitivity could be improved by fine tuning the FMR position. The sensitivity remains around 1 nT Hz$^{-1/2}$ within the linewidth of the mechanical resonances (about 10 MHz), for five mechanical modes. 

\begin{figure}[t!]
 \centering
 \includegraphics[width=8.9cm]{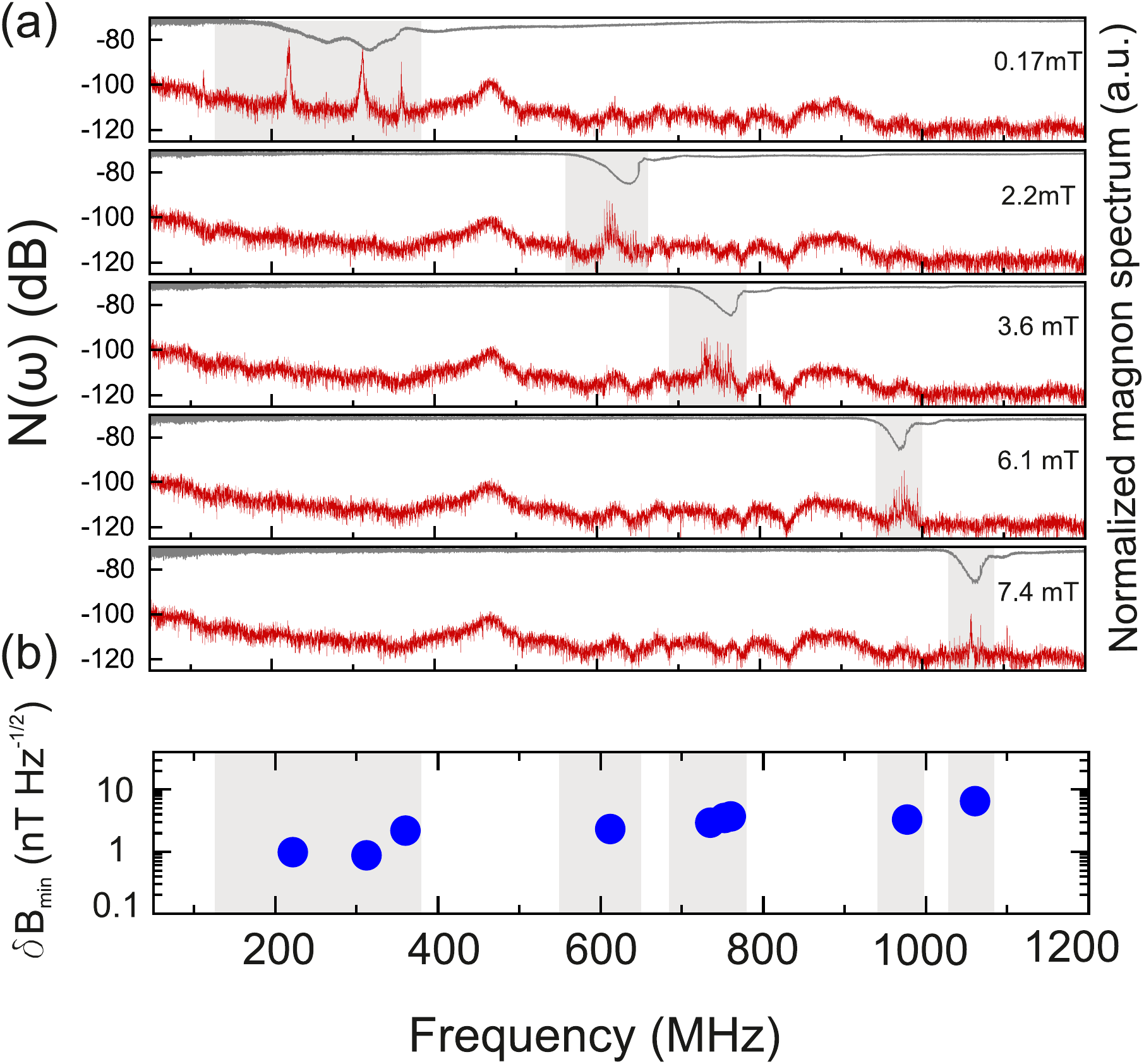}
      \caption{ (\textbf{a}) System response  for different FMR modes excited on the YIG film. Gray curves show magnon resonances for different values of static magnetic field. (\textbf{b}) Peak magnetic field sensitivity obtained by moving the FMR mode.}
\label{fig8}
\end{figure}

In Fig. \ref{fig8}a, we report the system response $N(\omega)$ as a function of frequency for different values of $B_{DC}$. We use a second sphere with a similar radius, mechanical spectrum and optical quality factors. As evidenced by the FMR spectra in Fig. \ref{fig8}a (grey curves), the FMR frequency increases increasing $B_{DC}$, and shifts the magnetomoeter spectral response $N(\omega)$ (red curves). In addition, this shift is acompanied by a spectral narrowing of the OMM bandwith, clearly following a spectral narrowing of the FMR dip (Fig. \ref{fig1}f). This behavior confirms that the magnetic signal appears only where the linewidth of the FMR resonance overlaps with the mechanical resonances of the microsphere. It also evidences the presence of mechanical modes that are hidden below the noise level of the SA. In Fig. \ref{fig8}b we obtain the peak sensitivity for the different measured positions of the FMR mode. We note that the OMM detects magnetic fields even above an operational frequency of 1 GHz (see Supplementary Information). The measured sensitivities are comparable with the one reported in Fig. \ref{fig8}d ($\thicksim$ 1 nT Hz$^{-1/2}$). This value of the operational frequency is a lower bound limited by the minimum $B_{DC}$ reachable in our experimental set up. 

As noted above, the FMR resonances obtained are rather broad given that the electromagnets were placed several centimetres away from the sample, resulting in a not fully uniform B$_{DC}$. The linewidths of the FMR resonances shown in  Fig. \ref{fig6}c are a factor of two larger than the  one shown in Fig. \ref{fig1}f, which were obtained measuring the YIG film in a set up with a highly homogeneous B$_{DC}$. The origin of this broadening is due to the excitation of non-uniform modes and low-field losses \cite{nembach2011,schloemann1992}. On the positive hand, under these conditions, the operational frequency range is increased and several mechanical modes can be covered without changing $B_{DC}$. On the other hand, a much improved sensitivity could be attained with a narrower and deeper FMR resonant with a mechanical mode. 

A further evidence that the mechanical modes of the spheres are excited by mechanical vibrations within the YIG layer generated by magnetostriction is given by additional experiments performed with a high-frequency Doppler Vibrometer. This setup implements an optical technique for non-contact measurements of displacement in the vertical direction with picometer accuracy. We use this technique to measure the YIG surface displacement without the microsphere. The measured displacement as a function of frequency results in non zero amplitudes only in a frequency range that is coincident with a given FMR (See the Supplementary Information). The deformation spatial profile is in phase throughout the YIG layer surface, i.e., there is a spatially homogenous out-of-plane displacement. We cannot rule out the excitation of phonon modes with in-plane deformation within the YIG layer, but those are not playing an active role activating the mechanical modes of the sphere, which is also verified with Finite Element Method (FEM) simulations.

The field sensitivity presented in Fig. \ref{fig6}e, is similar to the best sensitivity obtained in previous cavity OMM studies \cite{bowen,onchipmagnetometer,quantummag,fostneradv}. In those references, a magnetostrictive material (Terfenol-D) was used due to its high magnetostrictive coefficient \cite{Verhoeven}. Despite the fact that single crystal YIG was found to be around a factor of two less magnetostrictive than Terfenol-D \cite{YIGmagnetostriction}, the high performance of the OMM reported here is due to the use of YIG to display a high quality FMR and a high Q glass resonator.

Compared with room temperature devices like diamond NV-centers magnetometers, the device reported here shows a factor of two smaller peak sensitivity than the sub-pico Tesla NV magnetometer reported in Refs. \cite{casola2018probing, esr2008}, with the advantage of having a fiber-based optical detection. The sensitivity values reported here outperform electrical Lorentz-Force magnetometers \cite{Givens1999} of comparable size by three orders of magnitude. SQUID magnetometry can detect magnetic fields that are five orders of magnitude smaller than our scheme \cite{kirtley1995,Prance2003}, reaching sensitivities of 1 fT Hz$^{-1/2}$ at $\thicksim$ 100 Hz, but it requires cryogenic environments to operate. 


In summary, we have demonstrated a hybrid system composed of a magnetic resonator coupled by mechanical interaction to a whispering gallery mode optomechanical cavity to detect weak oscillating magnetic fields. A peak magnetic field sensitivity of $\sim$ 850 pT Hz$^{-1/2}$ is achieved by exciting a mechanical mode at 206 MHz. This value can be further improved by optimizing the overlap between the FMR resonance and the mechanical resonance of the optomechanical cavity. Besides the excellent figures of merit, the tuneability of the frequency response up to 1 GHz, room temperature operation and simplicity in fabrication offer the opportunity of designing a high-performance magnetometer. Large bandwidths are necessary for applications such as high-speed detection, mechanical signal processing and for high-resolution imaging methods \cite{bookmagnetometers}. In this regard, the frequency response of our device could be further extended to higher frequencies by increasing the static magnetic field. The magnetometer's sensitivity can be further improved following different strategies. For example, measuring at low temperatures or high vacuum conditions would result in better sensitivity values, since the sensitivity behaves as $\delta B_{min}(\omega)\sim \sqrt{T Q_m}$. Moreover, using a harder material than PMMA would reduce the mechanical impedance mismatch between PMMA and YIG, avoiding mechanical energy to be dissipated at the interface before reaching the sphere. In addition to the technological possibilities of designing a new magnetometer our hybrid device also opens a path towards studying phenomena related to phonon-magnon coupling. Currently, magnons are gathering increasing attention in spintronics experiments (e.g. magnonics \cite{Chumak2015} and spin caloritronics areas \cite{harii2019spin,costache2012magnon,Bauer2012}) as means of processing spin information and managing heat in nanoscale structures. Even though its superior properties make YIG a common choice for spintronic applications, the underlying physical mechanisms involved in phonon-magnon coupling are only analyzed by controlling the magnonic system. In contrast, our hybrid-resonator can be used as a novel approach to study phonon-magnon coupling, controlling the phonon contribution using optical techniques.

This work was supported by the Spanish Severo Ochoa Excellence program. D. N. U. gratefully acknowledges the support of a Ramón y Cajal postdoctoral fellowship (RYC-2014-15392) and the Ministry of Science, Innovation and Universities (PGC2018-094490-B-C22). M. V. C. acknowledges support from the Spanish Ministry of Economy and Competitiveness, MINECO (under Contract FIS2015-62641-ERC). M. V. C. thanks H. Tang and G. E. W. Bauer for discussions. M. F. C. acknowledges support of a Severo Ochoa studenship. G. A. aknowledges support of a BIST studentship. The authors thank G. Jakob for the YIG samples and M. Cecchini for experimental support.

\end{document}